# Deep Dynamic Neural Network to trade-off between Accuracy and Diversity in a News Recommender System

Shaina Raza[1] and Chen Ding[1]

[1] Ryerson University, Toronto, ON, Canada
shaina.raza@ryerson.ca, cding@ryerson.ca

**Abstract.** The news recommender systems are marked by a few unique challenges specific to the news domain. These challenges emerge from rapidly evolving readers' interests over dynamically generated news items that continuously change over time. News reading is also driven by a blend of a reader's long-term and short-term interests. In addition, diversity is required in a news recommender system, not only to keep the reader engaged in the reading process but to get them exposed to different views and opinions. In this paper, we propose a deep neural network that jointly learns informative news and readers' interests into a unified framework. We learn the news representation (features) from the headlines, snippets (body) and taxonomy (category, subcategory) of news. We learn a reader's long-term interests from the reader's click history, short-term interests from the recent clicks via LSTMSs and the diversified reader's interests through the attention mechanism. We also apply different levels of attention to our model. We conduct extensive experiments on two news datasets to demonstrate the effectiveness of our approach.

## 1 Introduction

Nowadays, there are many online web sites, where the news gets continuously published, which can cause the information overload problem for their readers. The news recommender system (NRS) can narrow down the limitless options and provide readers with the news based on what they have liked in the past. To understand its readers, an NRS needs to analyze the news content and make recommendations in real time.

For the news recommendation, there are three keys issues that need to be addressed. First, we observe that readers' interests are not fixed and tend to change over time. Generally, their interests on different topics are stable in the long run, while the content in which they are currently interested is often affected by their up-to-date concerns, or by certain events and contexts [1], such a breaking news, weather alerts. For example, a reader who is a fan of 'Cristiano Ronaldo' may have read many soccer related news for several years. This reflects the reader's long-term interests. Recently, impacted by the pandemic, the reader also began to browse news related to the COVID-19. This reflects the reader's short-term interests.

Second, we observe that the readers' interests naturally form a sequence over time. Each sequence consists of a list of the consumed items together with their associated timestamps. For example, we consider a reader who reads about 'Lionel Messi' at time $t_1$, 'Soccer tournament postponed due to COVID-19' at $t_2$, and then 'Is it safe to play soccer again' at $t_3$. In each time step, the reader's next click sequentially depends on the previous ones. In this paper, we use click as a general term representing any interaction (browse, click, comment, read) between the reader and the news item.

Third, we observe that readers may get bored of only reading news on similar topics. It is not wise to keep recommending similar news to a reader repeatedly. For example, the reader reads about the tennis player 'Novak Djokovic'. Based on the reader's interests, the NRS recommends him more similar news about Djokovic. After a while, the reader gets bored. Probably, he wants to read about other tennis players or different news. This hints towards recommending the diversified news (in addition to the personalized news) to the readers. We also observe that the short-term interests of the readers implicitly show the diversified patterns [2,3]. For example, while reading the sport-related news as usual, the reader might start browsing the news related to real estate market, due to his latest interest in purchasing a property. Or the reader is compelled to read US political news after getting the breaking news about the violent attack on the US Capitol in early 2021. In each such case, the diversified reading patterns are seen in the short-term interests.

To address these issues, the NRS needs to consider both the long-term and short-term interests of the readers. In addition, the NRS should capture the sequential patterns in the readers' clicks and recommend diverse news to the readers. This whole phenomenon in which the readers' interests change over time (for various reasons) is called the temporal dynamics [1].

Generally, based on the optimization objective [4], the recommendation approaches are categorized into two types: (i) the accuracy-based (aka relevance-based) [5] and; (ii) the diversity-based [6,7]. The accuracy-based approach optimizes the recommendation results by matching them with the user profile. A major limitation of this approach is that only the items that match with what the user has liked in the past are recommended. The diversity-based approach, on the other hand, aims at recommending a wide variety of items to the user no matter whether they are similar to the previously liked items.

In the state-of-the-art NRS [2,8], many of them apply the accuracy-based approaches, where the readers' long- and short-term interests are addressed while the diversity is often not considered. There is limited work [3,9] in NRS that considers the diversity. However, they have their own limitations: (i) the readers' interests are often taken as static; and (ii) a balance between the accuracy and diversity is not fully explored. Ignoring these factors could result in sub-optimal recommendations.

The research shows that there always exists a tradeoff between accuracy and diversity [10]. Not considering this tradeoff in an NRS has the following negative implications: (i) the recommendations based solely on accuracy could fail to suggest a variety of news items to the readers. The readers may get bored of reading similar news. They may also be trapped in echo chambers, where they get only the news that reflects the like-minded opinions; and (ii) conversely, the recommendations based only on diversity could potentially present news to the readers that totally deviates from their interests. The readers may leave the system after getting all the news that they are not interested in. To provide a better user experience, an NRS should extend beyond the conventional accuracy evaluation criteria and consider a balanced combined objective.

In this work, we aim to address the temporal dynamics issue and offer balanced news recommendations to the news readers. We still focus on accuracy but making sure the diversity is acceptable (at least the medium-level). We propose a deep dynamic neural network (**D2NN**) to provide effective news recommendations to the readers. Our proposed model has three components: (i) the news modeling; (ii) the reader interest modeling; and (iii) the recommendation part.

We design a novel news modeling component with BERT [11]. Convolution Neural Network (CNN) [12] and the attention mechanisms [13]. For each news item, we not only consider the news title and/or topics, similar to some recent work [2,8,14], but also consider various types of side information such as news body, taxonomy, timestamps, etc. The embeddings are obtained from the BERT model that considers the side information of the news data. Different from [2,8,14,15], we pretrain the BERT model on the news specific data to generate the embeddings. We use CNN on top of the embeddings to capture the local context information. We apply the word-level attention on the news to select the important words. We also apply the news-level attention to weigh each type of the side information according to its effectiveness in the NRS.

We design a novel reader modeling component to profile the reader's interests. We learn the reader's long-term interests from the whole history, the reader's short-term interests using a Long short-term memory (LSTM) network [5] on the recent history, and the diverse patterns in the reader's interests within a session (short-term interests) through an attention module. The intuition is that the reader's sequences consist of repeated patterns. With the LSTM, the reader modeling component can forget or retain some portions of the history. Using attention, the model learns the important clicks at successive steps and reduces the effect of repeating patterns. Finally, the recommendation module defines the probability of a candidate news article being clicked by a reader.

To the best of our knowledge, we are the first to provide such a wide list of aspects (features) related to the news (i.e., the side information) and readers in an NRS. There are three main contributions in our work:

(1) We address the dynamics in a reader's interests over time with seamless integration of long-term, short-term, diverse, and sequential interests. Our approach inherently provides a balance between highly accurate and reasonably diverse news recommendations.

(2) We emphasize the inclusion of heterogenous side information including the news headlines, snippets (passages from the news body) and taxonomy (category, subcategory) to learn the multifaceted news representations, consisting of textual, temporal, and contextual features from a sequences of news articles. The embeddings are generated from the BERT model that is pretrained on the news specific data. Most NRS [2,8,14] only consider the title to represent the news. There are also NRS that consider the news body, such as [15]. However, in [15], the embeddings (i.e., GloVe[1]) are pretrained on generic corpuses such as Wikipedia 2014. In contrast, we train the word embeddings on a news-specific corpus.

(3) Through our extensive experiments on two news datasets, we show that our proposed model has achieved a high-level of the recommendation accuracy meanwhile maintaining a requisite level of diversity. We also introduce a new metric to measure the tradeoff performance with respect to accuracy and diversity.

The rest of the paper is organized as follows. Section 2 is the related work. Section 3 discusses our D2NN framework. Section 4 explains the experiment setup. Section 5 analyzes the results. Section 6 gives the conclusion and lists the future work.

## 2  Related Work

***Temporal Dynamics:*** In a recommender system, the temporal dynamics refers to changing users' preferences over time [1]. The earlier work on temporal dynamics is based on simple time-decay functions [16]. A time decay function assigns different weights to a user's feedback and gives more importance to the recent interactions (or ratings) over the older ones. A general limitation of this method is that it emphasizes too much on the recent user feedback where the importance of the past feedback is underestimated. The past feedback is also important, in a recommender system, to determine the long-term user's preferences (interests).

Another classical method for temporal dynamics is the time binning method such as the timeSVD++ [17]. In binning-based methods, the longer time bins usually represent the users' long-term interests, while the shorter time bins usually reflect the users' short-term interests. A general limitation of this method is that the users' behaviors are strictly cut-off during different time bins [18], where each time bin is treated as a separate representative of users' behavior.

---

[1] https://nlp.stanford.edu/projects/glove/

Both these methods (time decay and time binning methods) treat the user-item interactions as static. In reality, a user's feedback forms a sequence in time, where the previous user' interactions usually depend on the user's next action or click. A sequential recommender considers the sequential dependencies between a user's past and his recent interactions to make a recommendation [19]. The session-based [20] is a special type of sequential recommender that makes the recommendations based on available information in the current session. In session-based methods, generally the items that a user has already consumed (during that session) or the items that have a high likelihood of being consumed next, are recommended. The classical methods for sequential prediction are the Markov methods [21]such as Factorizing Personalized Markov Chain (FPMC) and Hierarchical Representation Model (HRM). A general limitation of these methods is the growing number of states that are hard to manage. To solve this issue, a number of deep learning-based models have been introduced in the literature.

The RNN models have been employed to model the sequential dependencies in the recommender system [5]. They are also used for generating the session-based recommendations [22]. The well-known NRS that use the RNNs for sequential modeling are CHAMELEON [23] and LSTUR [2] models. The CHAMELEON [23] is a session-based NRS that uses RNNs to represent a reader's short-term interests. This model does not consider the readers' long-term interests. LSTUR [2] is another NRS that considers the reader' long-term and short-term interests. LSTUR uses the GRU network to represent a news reader's short-term interests, whereas the reader's long-term interests are represented from the whole click history.

The Reinforcement Learning (RL) is a sequential model, which is based on Markov Decision Process (MDP) [21]. In RL, an agent (e.g., a recommender system) interacts with the environment (e.g., news items, readers) over discrete time steps. The goal of the agent is to learn a policy that maximizes the accumulated rewards (recommendations). DRN [9] is a Deep Q-Network based RL framework for personalized online news recommendations. This method models the sequential interactions between the news items and the reader' feedbacks to achieve higher rewards (e.g. CTR). Generally, the RL methods suffer from the computational complexity issue due to ever increasing number of states.

Recently, the graph neural networks (GNN) have received much attention in the recommender systems [24]. These models exploit both the node attributes and graph structure to represent the users and items in a combined neighborhood space. For instance, DKN [8], a news recommender system, combines the knowledge graph (KG) with collaborative readers' feedbacks via a graph attention network. GNewsRec [25] is a another GNN-based NRS that constructs a reader-news-topic graph network to model the interactions between the readers and news.

Recently, the Transformer-based model [13] such as the BERT [11] has demonstrated remarkable performance in extracting the sequential relations from the text. It has been used in many domains, including the recommender systems. For example, the BERT4Rec [26] applies the BERT model to extract a sequence of user interactions in a downstream recommendation task. However, this model (BERT4Rec) does not have a separate user model and there is no consideration of explicit user/item contexts (side information). Also, the BERT model takes all the pieces of information as one long document, even though different types of side information may contribute differently in a model.

These above-mentioned state-of-the-art NRS [2,8,14,23,25] do represent the sequential, long-term and/or short-term interests of the news readers, however, a few limitations are noted. First, these models normally consider the news IDs or news titles to represent the whole news. However, there are also other pieces of information that may be more descriptive (e.g., the news story) or that reflect a reader's short-term or long-term preferences (e.g., topics, categories). Second these methods do not consider the variations in the readers interactions. The focus is mainly on the prediction accuracy. As a result, the recommended items are usually very similar with each other and may cover only a small fraction of items, with no consideration of diversity.

*Diversity*: Maximal marginal relevance (MRR) is a classical technique to increase the diversity of documents retrieved against a query in an information retrieval (IR) system. MMR tries to reduce the redundancy of retrieval results while preserving the relevance of the query results for documents. Basically, this technique evaluates the query results and provide another list of documents that are re-ranked. The same idea is also used in the recommender systems to include diversity during the re-ranking of recommended items [6].

Generally, the diversity is incorporated in the recommender systems in two ways: (i) as a two-stage recommendation strategy; and (ii) in the optimization model. In a two-stage recommendation strategy [6,27], an existing CF method is typically used to predict the missing ratings in the first stage, and the second stage is used to promote desired diversification through a modified ranking strategy (rather than conventional ranking scheme).

The diversity is these methods is usually evaluated in two ways: (i) the individual diversity [6] or (ii) the aggregate diversity [28]. Individual Diversity is quantified as the pairwise dissimilarity between items in a (given) user's recommendation list. This technique is used to assess the diversity from the user's perspective. The aggregate diversity, on the other hand, is measured as the number of unique items recommended across users. This scheme captures the system-centric notion of diversity.

The two-stage recommendation strategy usually makes use of a tuning parameter that is defined explicitly to control the tradeoff between the accuracy and the diversity. The selection of the parameter is based on heuristics, where the items rated above the threshold are considered as prospective candidates for recommendation. This approach, in addition to increasing the computational burden, does not guarantee an optimum solution.

In the optimization model, a weighted combination of diversity and accuracy is incorporated in a joint optimization strategy. Some work [29,30] consider diversity in the recommendation process through the use regularization terms on items' feature space. In a recent NRS [31], the regularization terms (through a unique combination of Lasso and Ridge regressions) are used to tradeoff between a high-level of accuracy and a reasonable amount of diversity.

Some other works, related to the optimization technique, use the Determinantal point processes (DPPs) [7] to introduce diversity among the set of items. DPPs are probabilistic models and can be used to address the balance between the accuracy and the diversity within a set of diverse items. However, the applicability of DPPs is subject to high complexity matrix operations on large datasets.

Dueling Bandit Gradient Descent (DBGD) [21] is another optimization technique to introduce diversity in the recommendation models. The DBGD is an online learning-to-rank algorithm based on multi-arm bandit algorithms and is used to model the exploration-versus-exploitation trade-off for the relative feedback. DBGD is recently used in a state-of-the-art NRS [9] to improve the recommendation diversity. Despite the robust design introduced in this model [9], a few things make it harder to implement this model in the real-world setting. First, the learning efficiency of this model is limited in a high-dimensional parameter space. Second, this method assumes only the binary feedback because there is no way of directly observing the reward of users' actions.

In this paper, we consider the short-term and long-term interests of the news readers. In addition, we consider the sequential dependencies among the readers' feedbacks and also address the diversified readers' interests from their recent feedbacks. We also consider rich side information from the news content to include in our recommendation model. Different from the previous work, we provide a balance between highly accurate yet reasonable diverse news recommendations.

## 3 D2NN FRAMEWORK

### 3.1 Problem Definition

Given a set of readers, a set of news items, and an interaction sequence created in the chronological order for each reader who has interacted with the news at a timestamp, the recommendation task is to predict the news item that the reader will interact next.

### 3.2 Notations

The notations used in this paper are given in Table 1:

**Table 1. Notations used in the paper**

| Notation | Description |
|---|---|
| *(i) R; (ii) N; (iii) CN; (iv) H; (v) S; (vi) T; (vii) $WE^h$; (viii) $E^{tc}$ and $E^{tsc}$; (ix) $CE^h$; (x) $Y^r$, $YE^r$,* <br> * All of the above are in a set format, e.g., $R = \{r_1, r_2, .., r_k\}$, $k; l; m; n_r$ are lengths of vector | Sets of *(i)* readers (*r*); *(ii)* news (*n*); *(iii)* candidate news (*cn*); *(iv)* headlines (*h*); *(v)* snippets (*s*); *(vi)* taxonomy (*t*) (*tc:* category; *tsc:* subcategory); *(vii)* word embedding (*we*) of *h* (same for *s*); (viii) embeddings (*e*) of *tc* and *tsc*; *(ix)* contextualized embeddings (*ce*) of *h* (also *s*); *(x)* reader's click history ($y^r$), embedding of reader's click history ($ye^r$) |
| *CNN: (i)* $\odot$*;(ii)* $WE^h$ *; (iii)* $b_h$*; (iv)* $N_f$*; (v)* $2K + 1$ *(vi) Tanh, Relu;* | *(i)* convolution operator; *(ii)* kernel *(iii)* bias of *h*; *(iv)* number of filters*; (v)* filter size*; (vi)* Activation Functions |
| $\tilde{h}$ ; $\tilde{s}$ ; $\widetilde{tc}$ ; $\widetilde{tsc}$; $\tilde{n}$; $\widetilde{cn}$; $\tilde{r}$; $\widetilde{r_{lt}}$; $\widetilde{r_{st}}$; $\widetilde{r_{st}^d}$; $\widetilde{y^r}$ | Representations of *h; s; tc; tsc; n; cn; r; r*'s long-term (*lt*) interests; short-term (*st*) interests; diversified interests (*d*); $y^r$ respectively |
| *Attention: (i)* $\alpha_i^h$*; (ii)* $\mu_i^h$*; (iii) V, v* (e.g., $V_h, v_h$); *(iv) q* (e.g., $q_h$); *(v)* $\alpha$ *(e.g.,* $\alpha_h$) | *(i)* attention weight of *i*th word in *h;(ii)* hidden representation of $ce_i^h$; *(iii)* projection parameters;*(iv)* query vector*; (v)* attention weight (e.g., of *h*) |
| *LSTM: (i) t (subscript); (ii)* $\sigma$*; (iii)* $\circ$ *; (iv)* $f_t$, $i_t$, $o_t$ *;(v)* $hi_t$ ; *(vi)* $\widetilde{ce}_t$; *(vii)* $ce_t$ ; *(viii) W* | *(i)* time step; *(ii)*sigmoid; *(iii)*item-wise product;*(iv)* forget, input, output gates;*(v)* hidden state;*(vi)* cell input;*(vii)* cell state; *(viii)* weight matrix |
| *Training:(i)* $\bar{\mathcal{X}} = \{\bar{x}_1, \bar{x}_2, .., \bar{x}_{i-1}\}, \bar{x}_i, \bar{y}\}$; *(ii)* $\bar{x}_j$; *(iii)* $\bar{x}_i$; *(iv)* $\bar{y}$, *(v)* $\rho$ *(vi)* $\mathcal{S} +$ *(vii)* $\mathcal{S} -$ | *(i)* training sample; *(ii) j*th clicked news; *(iii) i*th candidate news; *(iv)* label; *(v)* probability; *(vi)* positive sample; *(vii)* negative sample |

### 3.3 Model Architecture

Our proposed model D2NN has three components: (i) the news modelling (NM) (Fig. 1), (ii) the reader interest modeling (RIM), and (iii) the news recommendation (NR) components (Fig. 2). We first explain the NM, then the RIM and finally the NR component.

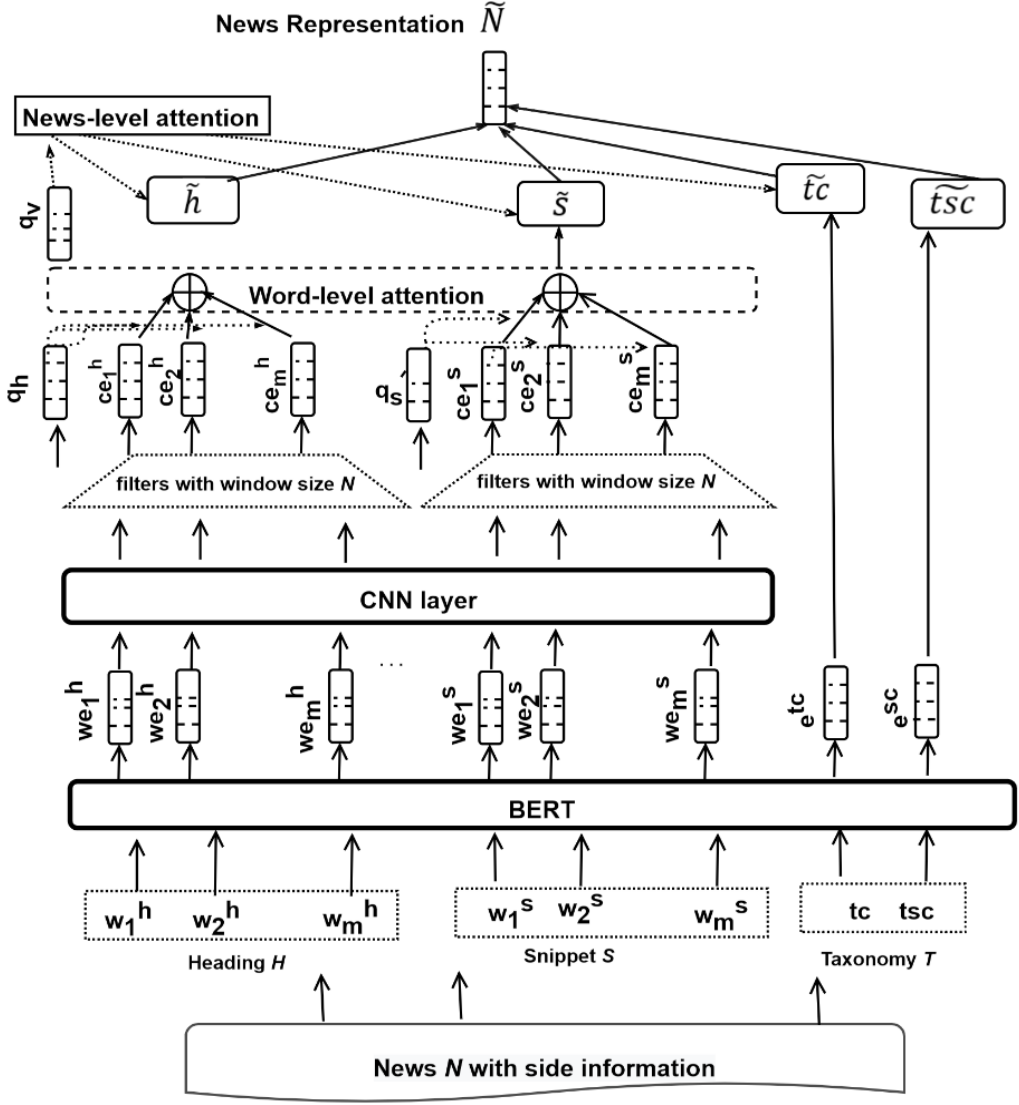

**Fig. 1.** News Modelling component

**News Modeling (NM) component:** The NM architecture is shown in Fig. 1 and discussed: below:

*Input and the side information:* The input to the NM is the news $N$. We give the heterogenous side information (i.e., headline, snippet, category, subcategory) from the news data as inputs to produce a combined news representation. The NM consists of three parallel neural networks (NNs), which respectively take inputs from the headline $H$, snippet $S$ and taxonomy $T$ to learn the news representations. There are three layers in NNs for $H$ and $S$, and two layers for $T$.

*BERT Embedding Layer:* The first layer in NNs for $H$ and $S$ is the embedding layer. The input to this layer is the words sequence from $H$ and $S$, and the output is the sequence of dense word embedding vectors $WE^h$ and $WE^s$ respectively. We utilize the token-level embeddings from the BERT model to generate the word embeddings. The advantage of our approach compared to other pretrained embeddings (e.g., GloVe or original BERT) is that we use the news specific data to generate the embeddings. In our preliminary experiments, we used a variety of methods to obtain the word embeddings. We tried the word embeddings pretrained on generic corpus (e.g., GloVe, original

BERT). We also pretrained the BERT model on the recent Wiki dumps[2]. In addition, we pretrained the BERT on the WikiNews[3] dumps and also on our news datasets. After all these efforts, we found that with the BERT trained on news specific corpuses, the performance of news recommendation methods is greatly improved.

The first layer in the NN for $T$ is also the embedding layer that converts the category $tc$ and subcategory $tsc$ from $T$ into the embeddings $E^{tc}$ and $E^{tsc}$ respectively.

*CNN layer:* The second layer in NNs for $H$ and $S$ is the CNN layer, to capture the local contexts (order of words) within $H$ and $S$. The local contexts are important to learn useful news representations. For example, in the headline "City prepares for the second COVID-19 wave", the local contexts "second" and "wave" are important to infer that this headline belongs to "COVID-19" event. We apply the CNN with different filters on the input word embeddings $WE^h$. The output from the CNN layer is the set of contextualized word embeddings $CE^h$, as shown in Equation 1.

$$ce_i^h = \text{ReLU}(WE^h \odot we^h_{(i-K):(i+K)} + b_h) \tag{1}$$

Here $\odot$ is convolution operator, $we^h_{(i-K):(i+K)}$ is the concatenation of word embeddings from position $(i-K):(i+K)$, $WE^h$ is the kernel and $b^h$ is the bias parameter. The parameters $N_f$ is the number of filters and $2K+1$ is the filter size. The CNN filter on $S$ is applied in the same way. We apply a separate filter on the snippet's word embeddings $WE^s$ to produce the context embeddings $CE^s$, as shown in equation 2.

$$ce_i^s = \text{ReLU}(WE^s \odot we^s_{(i-K):(i+K)} + b^s) \tag{2}$$

There is no CNN layer in the NN for $T$.

*Attention mechanism for the words:* The third layer in NNs for $H$ and $S$ is the attention layer. We apply the word-level attention to learn the related and relevant words from the news. In the headline "Thousands of people died of this COVID-19", the word "thousands" is linked to "people" and the word "this" is related to 'COVID-19'. Without the attention mechanism, the less significant words (e.g., "thousands" or "this") could easily be ignored. The input to the attention layer is the contextualized representations $CE^h$ and $CE^s$ from the CNN layer, and the output is the representations of the headline $\tilde{h}$ and snippet $\tilde{s}$.

We apply the attention on the contextualized representations of $h$, i.e., $CE^h$ to produce the headline representation $\tilde{h}$, as in Equation 3.

$$\tilde{h} = \sum_{i=1}^{M} \alpha_i^h ce_i^h \tag{3}$$

Here, $\tilde{h}$ is the summation of the contextualized representations ($ce_i^h$) of all the words in $h$ weighed by the attention weights $(\alpha_i^h)$. The attention weight $\alpha$ of the $i^{th}$ word in $h$ is given in Equation 4.

$$\alpha_i^h = \frac{\exp(\mu_i^h \mu_h)}{\sum_{j=1}^{M} \exp(\mu_j^h \mu_h)}, \quad \text{where } \mu_i^h = q_h \tanh(V_h ce_i^h + v_h) \tag{4}$$

Here $\mu_i^h$ is the hidden representation of $ce_i^h$, $\mu_h$ assists in deciding important words in $h$. $V_h$ and $v_h$ are the projection parameters and $q_h$ is the query vector of $h$.

The snippet representation $\tilde{s}$ is also calculated in the same manner as shown in Equation 5 and the attention weight $\alpha_j^s$ in shown in Equation 6.

$$\tilde{s} = \sum_{j=1}^{M} \alpha_j^s ce_j^s \tag{5}$$

---

[2] https://dumps.wikimedia.org/
[3] https://dumps.wikimedia.org/enwiki/

$$\alpha_i^s = \frac{\exp(\mu_i^s \mu_s)}{\sum_{j=1}^{M} \exp(\mu_j^s \mu_s)} \qquad \text{where } \mu_i^s = q_s^S \tanh(V_s^S ce_i^s + b_s) \tag{6}$$

There are limited words in taxonomy $T$, so there is no word-level attention for $T$. The second layer in the NN for $T$ is a dense layer that transforms the embeddings $E^{tc}$ and $E^{tsc}$ into the vector representation $\widetilde{tc}$ and $\widetilde{tsc}$, as shown in Equations 7 and 8 respectively.

$$\widetilde{tc} = ReLU(V_{tc} e^{tc} + v_{tc}) \tag{7}$$

$$\widetilde{tsc} = ReLU(V_{tsc} e^{tsc} + v_{tsc}) \tag{8}$$

*Attention mechanism for the news*: After building the NNs for *H, S* and *T*, we apply the news-level attention on all the learnt news representations. The intuition is that different side information has different usefulness and contributes differently. For example, if the snippet is less informative than the headline, then it should be weighed less, and the headline should be weighed more. So, we take each side information independently and weigh each piece of information accordingly. The final news representation $\tilde{n}$ is the summation of $\tilde{h}, \tilde{ts}, \widetilde{tc}, \widetilde{tsc}$, as in Equation 9.

$$\tilde{n} = \alpha_h \tilde{h} + \alpha_s \tilde{s} + \alpha_{tc} \widetilde{tc} + \alpha_{tsc} \widetilde{tsc} \tag{9}$$

The attention weight $\alpha_h$ is given in Equation 10:

$$\alpha_h = \frac{\exp(\mu^h \mu_h)}{\exp(\mu^h \mu_h) + \exp(\mu^s \mu_s) + \exp(\mu^{tc} \mu_{tc}) + \exp(\mu^{tsc} \mu_{tsc})} \tag{10}$$

Here $\mu^h = q_h \tanh(V_h \tilde{h} + v_h)$. The attention weights $\alpha_s, \alpha_{tc}, \alpha_{tsc}$ are calculated in a similar way.

The CNN and the attention layers is also found in the DKN [8], LSTUR [2] and NAML [15]. However, our network structure is different. Firstly, we include more side information (DKN [8], LSTUR [2] include the titles, topics). Though, NAML [15] includes news body, but the dataset in [15] is quite precise that includes one month user logs only. In contrast, one of our datasets (NYTimes) range over a longer period of time (two years data). This much data is good to model the temporal dynamics in readers' interests. Secondly, we generate the word embeddings by pretraining the BERT with the news specific data (they use embeddings pretrained on general data). And lastly, we consider the sequential readers' clicks (DKN considers the readers' current interests, whereas LSTUR and NAML considers readers' interests - without considering the diversified interests within sessions).

**Reader Interest Modeling (RIM):** The RIM is a sequential model that learns the readers' representations from the click history. We have two components in this module: the *long-term interest (LTI)* and the *short-term interest (STI)* modules. We also have a *diversity-aware interest modeling (DIM)* part within STI to introduce diversity in our model. The architecture of the RIM is shown in Fig. 2, with the following parts:

*Long-term interest (LTI) module:* The LTI captures the readers' long-term interests. It simply takes as input the whole sequential click history $Y^r$ of a reader and adds the embeddings of the reader's history in $YE^r$. The output from LTI is the sequential representation of the reader's long-term interests as $\widetilde{r_{lt}}$, as shown in Equation 11.

$$\widetilde{r_{lt}} = \sum_{i=1}^{N_r} YE_i^r \tag{11}$$

*Short-term interest (STI) module:* The STI captures the readers' short-term interests. It takes as input the reader's click history and outputs the short-term sequential representation $\widetilde{r_{st}}$. The output $\widetilde{r_{lt}}$ from LTI is sent to the LSTM network and the following steps take place that produce the short-term sequences, as shown in Equation 12.

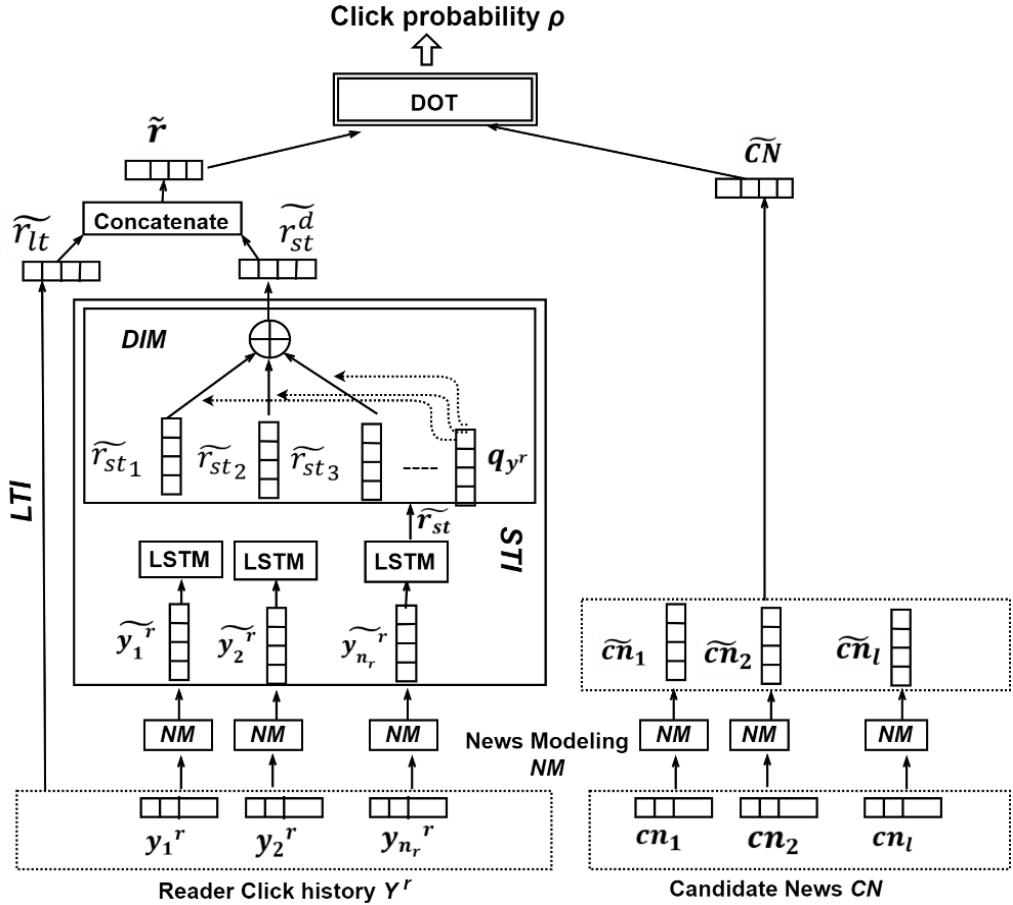

**Fig. 2.** Reader Modelling & recommendation component

$$\begin{align}
&\text{(i)} \quad f_t = \sigma\left(W_f \cdot [hi_{t-1}, \widetilde{r_{lt}}] + b_f\right), \\
&\text{(ii)} \quad i_t = \sigma\left(W_i \cdot [hi_{t-1}, \widetilde{r_{lt}}] + b_i\right), \\
&\text{(iii)} \quad \widetilde{ce}_t = \tanh\left(W_{ce} + [hi_{t-1}, \widetilde{r_{lt}}] + b_{ce}\right), \\
&\text{(iv)} \quad ce_t = f_t \circ ce_{t-1} + i_t \circ \widetilde{ce}_t, \\
&\text{(v)} \quad o_t = \sigma\left(W_o \cdot [hi_{t-1}, \widetilde{r_{lt}}] + b_o\right), \\
&\text{(vi)} \quad hi_t = o_t \circ \tanh(ce_t), \\
&\text{(vii)} \widetilde{r_{st}} = hi_t
\end{align} \tag{12}$$

The short-term representation is the last hidden state (vi) from LSTM, i.e., $\widetilde{r_{st}} = hi_t$.

***Diversity-aware interest modeling through attention:*** As hypothesized earlier, the diversity is intrinsically reflected in the readers' short-term interests. Generally, the readers' clicks among consecutive sessions are repeating, e.g., a reader reading about the COVID-19 in successive sessions. Due to these repeated clicks in sessions, the recommendations that are produced for these sessions are also similar. So, it is important to attend to readers' different clicks during different time steps for diversity.

The short-term sequences coming from the STI do not consider the variations in a reader's behavior. Inspired by the diversity-based attention in [32], we also use the attention to introduce diversity in our model. The diversity-based attention in [32] is applied on a content-based approach.

Different from [32], we apply the diversity for the user (reader) modeling task. We apply the attention on the short-term sequences $\widetilde{r_{st}}$ in the DIM part.

We denote the attention weight of the $i^{th}$ clicked news in $\widetilde{r_{st}}$ as $\alpha_i^{hi_t}$, which is calculated by evaluating the importance of interaction between the clicked news ($hi_t$) during time $t$ and the clicked news representation $\widetilde{r_{st}}$, as shown in Equation 13:

$$\alpha_i^{hi_t} = \frac{\exp(\mu_i^{hi_t}\mu_{hi_t})}{\sum_{j=1}^{|\widetilde{r_{st}}|}\exp(\mu_j^{hi_t}\mu_{hi_t})} \quad (13)$$

Here $\mu_i^{hi_t} = q_{hi_t}\tanh(V_{hi_t}\widetilde{r_{st}} + v_{hi_t})$. The final diversified representation $\widetilde{r_{st}^d}$ is the summation of the clicked news representations weighted by their attention weights, as shown in Equation 14:

$$\widetilde{r_{st}^d} = \sum_{i=1}^{|\widetilde{r_{st}}|} \alpha_i^{hi_t} \widetilde{r_{st}}_i \quad (24)$$

We combine the reader' long-term ($\widetilde{r_{lt}}$) and short-term diversified representations ($\widetilde{r_{st}^d}$) to produce the final reader representation $\tilde{r}$. By doing so, we provide a seamless integration of a reader's long-term and short-term diversified interests.

*News Recommendation (NR):* The NR component predicts the click probability of a reader based on the click history. We compute the click probability score $\rho$ of a reader $r$ clicking the candidate news $cn$, as $\rho(r, cn)$, by taking the dot product between their representations i.e., ($\tilde{r}, \widetilde{cn}$), as shown in Equation 15:

$$\rho(r, cn) = \tilde{r}^T \widetilde{cn} \quad (3)$$

### 3.4 Model Training

We use the negative sampling [2] to train the model. We train our model using positive (reader's observed clicks) and negative samples (unobserved history from the same session) with a ratio of 1:5. Each training sample $\bar{\mathcal{X}}$ has a clicked news, a candidate news and a label $\bar{y}$ ($\bar{y} = 1$ is positive, $\bar{y} = 0$ is negative), with an estimated probability $\rho \in [0,1]$ for each click. We minimize the loss for each sample using the negative log-likelihood function to train the model, shown in Equation 16.

$$\mathcal{L} = -\{\sum_{i=1}^{\mathcal{S}+} \bar{y} \log\rho_i + \sum_{i=1}^{\mathcal{S}-}(1-\bar{y})\log(1-\rho_i)\} \quad (46)$$

Here $\mathcal{S}+$ is positive and $\mathcal{S}-$ is negative sample set respectively. We take the reader's whole click history to train LTI. To train STI, we create time-ordered sessions (sequences) in the reader's click history, with both positive and negative samples. We use paddings to fill up the shorter readers' sequences. We also pad the headline and snippet inputs of shorter length.

## 4 Experimental Setup

### 4.1 Dataset and Experiment Settings

We evaluate our proposed model on the two real-world news datasets:

*New York Times (NYTimes):* we collected the news articles and the anonymized readers' interactions using NYTimes API. The interactions on articles were retrieved with respect to the timeline of the news data A sample of the dataset can be accessed here[4].

---

[4] https://github.com/deeplearningnrs/D2NN

12*MIcrosoft News Dataset (MIND):* we use a large-scale benchmark news dataset [33] consisting of anonymized behavior logs from the Microsoft News. It can be accessed here[5]. The original MIND dataset has about 1 million readers with 15 million clicks on 160k articles, and we use the MIND-small (a smaller version) in the experiments.

Both the datasets consist of English news articles. For both datasets, we generated the training samples from the click histories and impressions logs according to the format given in the MIND paper [33]. An impression log records the news articles that the reader visits or clicks at a specific time [33]. The basic statistics for both datasets are shown in Table 2.

Table 2. Basic statistics of the datasets

| Dataset | Duration | Readers | News | Clicks | News information | Reader information |
|---|---|---|---|---|---|---|
| **NYTimes** | 2 years (1st Jan. 2017 to 31st Dec. 2018) | 240,000 | 15,000 | 2,000,000 | ID, headline, snippet (abstract), category, sub-category, publication timestamp | ID, interaction timestamp, click history, impressions |
| **MIND - small** | 6 weeks (12th Oct. 2019 to 22nd Nov. 2019) | 50,000 | 161,013 | 156,925 | | |

### 4.2 Evaluation Methodology

We use the leave-one-out-evaluation for the next-item recommendation. We chronologically sort each reader's interactions and hold out the last item in the sequence as test set, second-last interaction as validation set and the rest as training set.

**Evaluation Metrics:** We evaluate all models for the following evaluation metrics:

*Root Mean Squared Error (RMSE):* to evaluate the prediction accuracy, we use RMSE [34]. Typically, RMSE are used in a recommender system to evaluate the difference between the predicted rating (estimating user' like or dislike) and the known rating that is given by the user.

*Area Under Curve (AUC):* as, the problem defined in this work is a click prediction problem, so we also use the AUC [34]. Typically, the AUC is used to calculate the area under the ROC (Receiver Operating Characteristic) curve for classification problems, and a higher value of AUC means that recommendations are better.

*Normalized Discounted Cumulative Gain (NDCG):* to measure the recommendation and ranking accuracy, we use NDCG @k [35]. Typically, an item is viewed as either relevant or not relevant by the other ranking measures (e.g., by precision, recall), while there can be degree of relevancy of items. NDCG uses the graded relevance to rank each item. Compared to F1-score [35] that is also commonly used to measure the recommendation accuracy, NDCG additionally considers the ranking order of the positive results (good recommendations) in the top *k* list. It is important for an NRS to place the good results in higher positions because such news items have better chances to be selected, especially when the readers do not have the patience to go through the whole *k* results. In this work, we focus on the NDCG to evaluate the ranking performance.

*Diversity (DIV):* to calculate the diversity, we compute the individual diversit as the average dissimilarity [6] of all the pairs of items in a user's recommended list at a specific cut-off value (k). In that, we calculate the recommendation diversity DIV @k using the intra-list similarity (ILS) [6,34].

---
[5] https://msnews.github.io/

*tradeoff*: we define a tradeoff metric that can measure the trade-off degree between accuracy (the mean NDCG) and diversity (mean DIV) as: *tradeoff=2*accuracy*diversity/ (accuracy+diversity)*. These mean scores are calculated over all *k* values.

We consider four *k* values: 5, 10, 20, and 50 for all metrics except RMSE and AUC. Our general assumption for choosing these evaluation metrics is that the readers will be more satisfied if the recommended results are accurate (matching their interests) and diverse (giving something new).

**Task Settings**: We pretrain the BERT on Wikinews<sup>Error! Bookmark not defined.</sup> dumps for the word embeddings. We implement our models using TensorFlow on the GPUs provided by Google Colab Pro[6]. For the hyperparameters, we consider:

- Dimension size *D* in range: {50, 100, 200, 300}
- Dropout rate *dr* in {0,0.1, ... ,0.9}
- $l_2$-regularization in range {1e-8, 1e-7, ..., 1e-1},
- CNN filter $C$ in {50, 100, 200, 300} with filter size *fs* in range {2, 3, 4}.
- The model is trained using Adam optimizer [39] with learning rate $\lambda$ is in the range {1e-5, 1e-4,...,1e-1}, and the parameters $\beta_1$, $\beta_2$ are set to default. The $\beta_1$ and $\beta_2$ are the Adam's configuration parameters that control the decay rates of the estimates.
- Batch size *bs* is considered in {64, 128, 256, 512}.

The optimal parameters are determined by the validation set as $D = 300$, $dr = 0.2$, $\lambda = 0.001$, $\beta_1 = 0.9$, $\beta_2 = 0.999$, $C = 200$, $fs = 3$, $bs = 256$, and $l_2 = 0.001$. For fair comparison, all baselines are configured to their optimal parameters. Each experiment is repeated 10 times.

### 4.3 Baselines

In our work, we want to compare our model with the baselines in terms of the accuracy (prediction accuracy, recommendation accuracy, ranking accuracy), diversity and most importantly tradeoff. The expectation (based on our claimed contribution) is that our tradeoff score should be the best, accuracy should also be high, and diversity should be reasonable. We use a mix of baselines in our experimentation and group them into three categories: (i) NRS, (ii) general DL models, (iii) sequential models as discussed below:

**NRS as baselines**: The first group of baselines is of NRS and consists of:

*DKN* : Deep Knowledge-Aware Network for News Recommendation [8] is a NRS that fuses the semantics and knowledge from the news for the click prediction problem.

*LSTUR*: A Neural News Recommendation Approach with both Long And Short-Term User Representations [2] is an NRS that learns the news and readers' (long-term, short-term) representations for the click prediction task. It has another variant **LSTUR-ini** with only short-term module.

**General Deep Learning Recommenders as baselines:** The second group is of general deep learning recommenders and consists of:

*DMF*: Deep Matrix Factorization Models for Recommender Systems [36] is a CF model that learns the similarities among the users and items to predicts the rankings of items.

*CDAE:* Collaborative Denoising Auto-Encoders for Top-N Recommender Systems [37] is another CF model that uses the idea of Denoising Auto-Encoders to model users' preference with implicit feedback.

---
[6] https://colab.research.google.com/signup

**Sequential Recommenders as baselines:** Our third group is of sequential (session-based) recommenders and consists of**:**
*GRU4Rec+* **:** A Gated Recurrent Unit for Recommendation with Improved Version [5] is a recommender system that uses RNNs to model the user-item interaction sequences for session-based task. Compared to simple GRU4Rec [22], GRU4Rec+ [5] adopts a different loss function and a different sampling strategy.
*SASRec* **:** Self-Attentive Sequential Recommendation [38] is a self-attention based sequential model for the task of next item recommendation. It captures the entire user sequence to predict the next item.
*SRGNN* : Session-Based Recommendation with Graph Neural Networks [39] aims to predict a user's next action based on the session information.
*BERT4Rec* : Sequential Recommendation with Bidirectional Encoder Representations from Transformer [26] adapts the BERT architecture based on the Transformer model for sequential recommendations. It captures the user's sequential behavior for the task of click prediction.

Among the baselines, DKN, LSTUR, SASRec, SRGNN and BERT4Rec are attention-based. We share the implementation details of our baselines also in this link[4].

## 5 Results and Analyses

### 5.1 Overall performance comparison

The comparison between our model and all the baselines is shown in Table 3 (bold is the best score). We discuss the results on both datasets (NYTimes and MIND), if there is a difference because of the dataset, we highlight the point explicitly.

Overall, we can see in Table 3 that our D2NN model has the highest prediction accuracy (the lowest RMSE) and the highest click prediction accuracy (the highest AUC). We also get the highest recommendation accuracy (highest NDCG for all the *k* values in NYTimes and the highest NDCG for *k* @20 and 50 in MIND). We achieve a reasonable level of diversity (medium-level DIV) for all the *k* values. The tradeoff score of our model is the highest among all models, which shows that we achieve the right balance between the high accuracy and reasonable diversity.

Overall, we get the better performance with the NYTimes. This is probably because our NYTimes dataset covers a longer range of readers' sequences and the news items. Our D2NN method is designed to model the sequential information from the news, and the readers' long-term and short-term interests in a better way. The MIND dataset is a short sequence dataset, which is better modeled by the recommenders focusing more on the short-term user modeling. However, despite the nature of both datasets, our D2NN model shows the best tradeoff score (also most of the accuracy measures) with both datasets.

Among the baselines, we find that the overall performance of the NRS baselines is better than that of the general methods. It can be seen with the higher tradeoff scores of DKN and LSTUR compared to other baselines. This is because these NRS methods were designed from the beginning to learn the news and reader representations, while in general methods, we need to explicitly provide this information.

Table 3. RMSE, AUC, NDCG @k, DIV @k and tradeoff scores of D2NN and the baselines

| Metric/Model | NRS | | | General Methods | | Sequential recommenders | | | | Our model |
|---|---|---|---|---|---|---|---|---|---|---|
| | DKN | LSTUR | LSTUR-ini | DMF | CDAE | GRU4Rec+ | SASRec | SRGNN | BERT4Rec | D2NN |
| *NYTimes Dataset* | | | | | | | | | | |
| RMSE | 0.573 | 0.897 | 0.807 | 0.671 | 0.667 | 0.649 | 0.641 | 0.591 | 0.529 | **0.506** |
| AUC | 0.544 | 0.500 | 0.511 | 0.489 | 0.404 | 0.215 | 0.360 | 0.478 | 0.505 | **0.590** |
| NDCG@5 | 0.278 | 0.150 | 0.180 | 0.255 | 0.199 | 0.045 | 0.127 | 0.141 | 0.079 | **0.387** |
| NDCG@10 | 0.297 | 0.212 | 0.222 | 0.247 | 0.204 | 0.047 | 0.138 | 0.140 | 0.098 | **0.449** |
| NDCG@20 | 0.310 | 0.307 | 0.307 | 0.270 | 0.288 | 0.055 | 0.156 | 0.157 | 0.123 | **0.511** |
| NDCG@50 | 0.439 | 0.333 | 0.300 | 0.339 | 0.299 | 0.081 | 0.212 | 0.197 | 0.167 | **0.567** |
| DIV@5 | 0.513 | 0.493 | 0.683 | 0.368 | 0.522 | **0.858** | 0.641 | 0.599 | 0.163 | 0.488 |
| DIV@10 | 0.495 | 0.536 | 0.726 | 0.383 | 0.302 | **0.836** | 0.620 | 0.585 | 0.235 | 0.526 |
| DIV@20 | 0.490 | 0.605 | 0.795 | 0.423 | 0.301 | **0.831** | 0.628 | 0.603 | 0.346 | 0.605 |
| DIV@50 | 0.428 | 0.638 | 0.828 | 0.485 | 0.313 | **0.838** | 0.668 | 0.647 | 0.415 | 0.670 |
| tradeoff | 0.392 | 0.348 | 0.379 | 0.333 | 0.279 | 0.107 | 0.254 | 0.252 | 0.166 | **0.521** |
| *Mind Dataset* | | | | | | | | | | |
| RMSE | 0.501 | 0.376 | 0.428 | 0.457 | 0.659 | 0.605 | 0.625 | 0.598 | 0.555 | **0.346** |
| AUC | 0.536 | 0.528 | 0.520 | 0.405 | 0.422 | 0.204 | 0.285 | 0.320 | 0.541 | **0.538** |
| NDCG@5 | **0.335** | 0.233 | 0.229 | 0.187 | 0.220 | 0.044 | 0.205 | 0.125 | 0.074 | 0.237 |
| NDCG@10 | **0.323** | 0.297 | 0.291 | 0.203 | 0.215 | 0.071 | 0.196 | 0.126 | 0.109 | 0.322 |
| NDCG@20 | 0.340 | 0.349 | 0.341 | 0.249 | 0.213 | 0.106 | 0.196 | 0.142 | 0.222 | **0.368** |
| NDCG@50 | 0.379 | 0.391 | 0.387 | 0.262 | 0.240 | 0.110 | 0.224 | 0.184 | 0.267 | **0.427** |
| DIV@5 | 0.528 | 0.328 | 0.401 | 0.342 | 0.391 | **0.818** | 0.604 | 0.714 | 0.101 | 0.447 |
| DIV@10 | 0.512 | 0.412 | 0.408 | 0.298 | 0.411 | **0.806** | 0.605 | 0.715 | 0.484 | 0.488 |
| DIV@20 | 0.409 | 0.425 | 0.411 | 0.312 | 0.417 | **0.806** | 0.608 | 0.720 | 0.409 | 0.481 |
| DIV@50 | 0.428 | 0.425 | 0.580 | 0.372 | 0.420 | **0.814** | 0.615 | 0.731 | 0.420 | 0.485 |
| tradeoff | 0.397 | 0.353 | 0.357 | 0.268 | 0.288 | 0.150 | 0.307 | 0.240 | 0.228 | **0.403** |

We also find that the general performance of the NRS methods is better than the CF models (DMF and CDAE). The CF suffers from the data sparsity problem that's why these methods have relatively lower recommendation accuracy. Also, the CF suffer from the inherent popularity bias, which explains why there is lower diversity. It indicates that we should go beyond the rating-only CF by adding content-based, contextual and attention-based solutions in an NRS.

We observe that the non-sequential models, in general, perform better than the sequential models, particularly on the accuracy and tradeoff scores. This is probably because a pure sequential recommender system usually takes the strictly ordered reader and item interactions as sequences. However, there is quite an uncertainty in the reading behaviors in an NRS that couldn't be captured by pure sequential models.

We also find that the overall performance of the attention-based methods is better than the traditional neural recommenders. This is seen with the relative better performance of the attention-based NRS methods compared to the general methods without attention. The accuracy of GRU4Rec+ (non-attention) is also lower among the sequential recommenders. This result indicates the effectiveness of the attention mechanism in focusing on details and selecting the important information for the readers.

Although the non-sequential recommenders are in general better than the sequential ones in terms of accuracy; but in terms of diversity, the sequential recommender GRU4Rec+ has the highest diversity among all the models on both datasets. These sequential models have the higher diversity on both datasets. It is important to mention here that the highest diversity does not mean the best performance. This higher diversity level is achieved with a big loss on accuracy. In an NRS, we would never want to achieve a high diversity at the cost of a big loss on accuracy [31]. In the subsequent results (Table 4), we find that the diversity of one of our model variants is comparable (very close) to the highest diversity of GRU4Rec+, which shows that by adjusting our model component, we can achieve a high diversity if that is the goal.

In the following experiments, we show the results on AUC and NDCG that are considered as standard evaluation metrics for recommendation accuracy [33]. We also show the DIV and the tradeoff score. The results of D2NN variants show similar patterns on both datasets, so we report the results on the NYTimes dataset only.

**5.2   Ablation Study**

We perform an ablation study to analyze the impact of different components in our model. Table 4 shows the performance of our default D2NN method and its variants. D2NN consists of LTI, STI, $s$ (snippet), $t$ (taxonomy) and $h$ (headline). Our model variants are named as: D2NN-LTI (D2NN with only LTI), D2NN-STI (D2NN with only STI). When we remove some side information, we use a minus sign as the superscript, e.g., D2NN-STI($s^-t^-$) is D2NN-STI without $s$ and $t$. We observe the similar patterns on NDCG and DIV for all the $k$ values, so we show the average of them (i.e., the mean NDCG and the mean DIV).

**Table 4.** AUC, mean NDCG, mean DIV and the tradeoff score of the D2NN variants

| Model/ Metric | AUC | NDCG | DIV | tradeoff |
|---|---|---|---|---|
| **D2NN** | **0.590** | **0.479** | 0.572 | **0.521** |
| D2NN ($s^-$) | 0.569 | 0.464 | 0.529 | 0.494 |
| D2NN ($s^-t^-$) | 0.550 | 0.387 | 0.519 | 0.444 |
| **D2NN-STI** | 0.530 | 0.292 | 0.729 | 0.416 |
| **D2NN-LTI** | 0.528 | 0.275 | 0.479 | 0.349 |
| D2NN-STI ($s^-$) | 0.501 | 0.275 | 0.713 | 0.397 |
| D2NN-STI ($s^-t^-$) | 0.566 | 0.297 | **0.817** | 0.436 |
| D2NN-LTI ($s^-$) | 0.480 | 0.265 | 0.462 | 0.337 |
| D2NN-LTI ($s^-t^-$) | 0.430 | 0.255 | 0.445 | 0.324 |
| D2NN ($w^-$) | 0.390 | 0.279 | 0.635 | 0.387 |
| D2NN ($n^-$) | 0.389 | 0.277 | 0.628 | 0.385 |
| D2NN ($r^-$) | 0.367 | 0.219 | 0.467 | 0.298 |

*Impact of Reader Interest Modeling*

We demonstrate the impact of modelling readers' long-term, short-term and combined interests in our D2NN model. As shown in Table 4, the overall accuracy and diversity scores of the D2NN-STI (modeling short-term and diverse interests) are higher than those of the D2NN-LTI (modeling only long-term interests). This is because in a typical news dataset, there are more short-term users, without considering short-term interests, the model performance drops.

We also see that our sequential model with LSTM and the attention outperforms the Transformer (sequence-to-sequence) model in modeling the readers' short-term interests. It is shown, in Table 4, by the better accuracy and diversity of our D2NN-STI compared to scores of BERT4Rec (in Table 3). The LSTM can model the readers' short-term interests in different shorter time ranges, whereas the attention can model the related and diversified readers' interests. Our model also has far less parameters. As, news reading is a sequential process (usually left to right) that is better conceived by RNNs, there is no need to consider bi-direction as in Transformers.

D2NN-STI has a built-in diversity-ware module DIM with the reader-level attention. We explore the impact of removing the reader-level attention from the model. The variant D2NN$(r^-)$ refers to D2NN without the reader-level attention. As shown in Table 4, the model accuracy drops when we remove the reader-level attention, as shown in D2NN$(r^-)$. As the primary purpose of the reader-level attention in STI is to include the diversity, removing attention from the STI obviously impacts the model diversity. We also see that diversity of D2NN$(r^-)$ is close to that of the D2NN-LTI. This is obvious as D2NN-LTI is by default without the reader-level attention. Overall, we see that it is necessary to include both STI and LTI, that's why the default D2NN performs the best.

*Impact of Side Information in News Modeling*

We demonstrate the efficacy of our D2NN model with respect to including the side information in the NM component. The results, in Table 4, show that the D2NN with all the side information is more competitive than the models that exclude one or more pieces of the side information. This is shown with the overall better accuracy of D2NN than D2NN$(s^-)$, which is better than D2NN$(s^-t^-)$.

We also demonstrate the impact of removing the side information in our model variants D2NN-LTI and D2NN-STI and the results are: (i) when we remove only the snippet from both variants, the overall accuracy of both models gets lower. This is obvious since we can better know about readers when we have more information about the clicked news; and (ii) when we remove the snippet and taxonomy from both variants, the performance of D2NN-STI increases and that of D2NN-LTI decreases. This shows that headlines alone are probably enough to make recommendations when the reader sequences are typically short. The lower performance in D2NN-LTI$(s^-t^-)$ is mainly because of skipping the news categories. The news taxonomy browsed by the reader has decisive influence on the long-term behavior. Finally, combining all the side information improves the performance as seen in the best performance of D2NN.

We also explore the impact of different levels of attention in the NM component. The D2NN$(w^-)$ refers to D2NN without the word-level attention, and D2NN$(n^-)$ is D2NN without the news-level attention. We see, in Table 4, that the model performance drops when we remove the word-level attention. This shows that word-level attention is important to capture the relatedness of the words. The performance drops (slightly) more when we remove the news-level attention. This shows the usefulness of news-level attention in learning different side information for the news representation. Overall, we see that both attention levels (word, news) are very important. This is demonstrated with the better scores of the default D2NN.

## 5.3 Effectiveness of the Self-Attention Mechanism

We study the effectiveness of including the self-attention in our model. We make two changes in the D2NN: (i) we replace the CNN layer in the NM with the self-attention; and (ii) we replace the attention in DIM with the multi-head self-attention. The goal is to see if we can perform better with multi-heads. We produce the contextualized representations of headline $h$ using self-attention, as in Equation 17.

$$ce_{i,k}^h = V_k^h \sum_{j=1}^{M} \alpha_{i,j}^k we_j \qquad (17)$$

Here $ce_{i,k}^h$ is the multi-head representation of $i^{th}$ word in the headline $h$ with $k^{th}$ attention head. The attention $\alpha_{i,j}^k$ is shown in Equation 18.

$$\alpha_{i,j}^k = \frac{\exp(we_i^H Q_k^h we_j)}{\sum_{m=1}^{M} \exp(we_i^H Q_k^h we_m)} \qquad (18)$$

Here $\alpha_{i,j}^k$ refers to the relative importance of the interaction between the $i^{th}$ and $j^{th}$ words in $k^{th}$ self-attention head. $V_k^h$ and $Q_k^h$ are the project parameters in the $k^{th}$ head and $ce_i^h$ is the summation of $\hbar$ attention heads i.e., $ce_i^h = [ce_{i,1}^h, ce_{i,2}^h, \ldots, ce_{i,\hbar}^h]$. The context representation for the snippet $s$ i.e., $ce_{i,k}^s$ is also calculated in the same manner. We also learn the reader's representation who clicked news in the same way.

Multi-head attention is particularly useful in scenarios when we want to learn different types of information from text [12]. For example, in the news "Ontario identify new COVID-19 cases, with a high number", the word 'COVID-19' here has interactions with many words i.e., 'new', 'cases' and 'number'. This kind of problem where we need to break the concept word by word, the multi-heads can be utilized to absorb knowledge from each word.

In this experiment for self-attention mechanism, we select a number of heads in the self-attention layer and report the results with few numbers i.e., 3, 5, 10 with maximum 16 heads (maximum. 16 is also seen in a previous work on multi-heads [40]). We compare these D3NN variants with our original D3NN that has vanilla (additive) attention. We show the effect of multi-heads on our model performance in in Fig. 3.

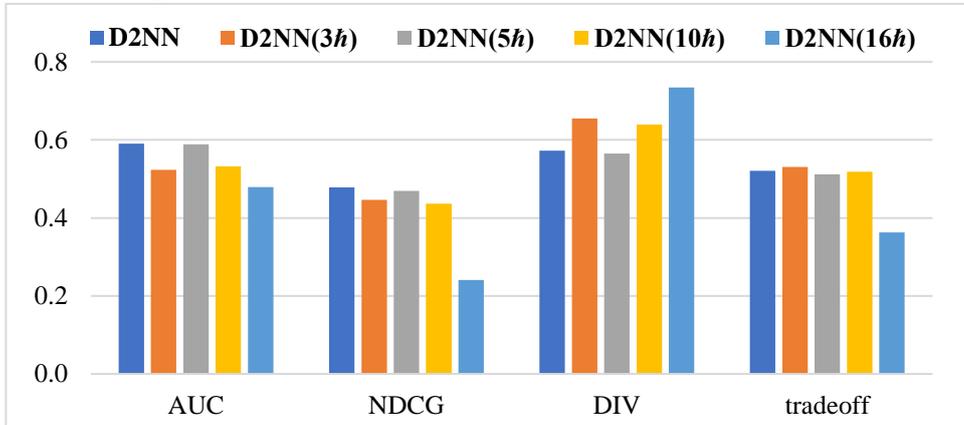

**Fig. 3.** D2NN with multi-head attention

The results, in Fig. 3, show that the default D2NN is better considering its overall performance, especially in terms of AUC, NDCG and tradeoff. The D2NN(3$\hbar$) shows marginal higher tradeoff than D2NN, which is probably related to the higher diversity. We also observe a trend with the increase or decrease in the number of heads in these variants. We find that as we increase the number of heads from 3$\hbar$ to 5$\hbar$, the model accuracy increases, but then the accuracy drops when we further increase the heads to 10$\hbar$ and 16$\hbar$. D2NN(16$\hbar$) has the lowest accuracy. In terms of diversity, we see the opposite trend to accuracy. The variants with lower accuracy now show the increased diversity. This indicates that there exists a negative correlation between the accuracy and the diversity.

Besides the recommendation tasks, we also see the effect of attention heads on model efficiency. Similar to the results reported in [40], we find the pruning half the heads could result in better model efficiency. In our experiment, we also find that the attention heads after a certain number become overhead for the model. Our original D2NN is not using multi-heads and we get a good performance.

### 5.4 Effectiveness of the Language Modeling

We use the BERT for the news language modeling (representation) task. We pre-train BERT on WikiNews[3] dumps (same timeline as our datasets), then we fine-tune BERT on our datasets. We use only the token-level embeddings from BERT. We compare our model using these language models (i) (original) BERT pretrained on Wikipedia and Books; (ii) our own trained BERT-News; (iii) the GloVe[1] model.

The results in Fig. 4 show the overall best performance of our model with the BERT-News. The performance of (original) BERT and the Glove models is relatively lower. This is probably because these models might have missed many domain-specific words in the embeddings. Th embeddings by these pretrained GloVe and (original) BERT models also contain some noise, that's why we see an unexpected rise in diversity. Overall, the result shows that the domain-specific language models are more useful for understanding news articles.

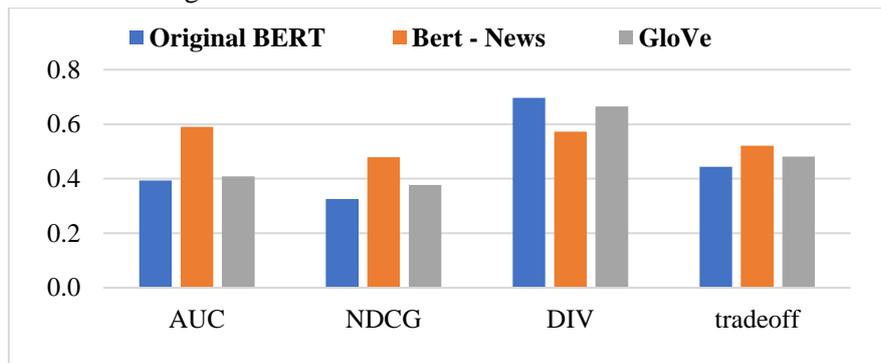

**Fig. 4.** Different news representations

### 5.5 Analyses on other hyperparameters

We also explore the impact of various important hyperparameters in the experiments. While analyzing one hyperparameter, the remaining hyperparameters are fixed at their optimal settings. Here, we report the broad findings by using various hyperparameters as discussed below:

*Negative: Positive samples:* We try different ratio of negative to positive samples in our model. The result shows that too few negative samples do not represent readers' interests sufficiently, while too many negative samples bring noise. We find that when we use too small number of negative samples (e.g., less than 5) for training, the model performance drops. This is because we are relying too much on positive implicit feedbacks (clicks) to produce similar recommendations with least consideration for other items. However, when we include too many negative samples, the model performance again drops. This is probably because our model begins including with too many negative samples which results in bringing noise in the model. Therefore, a ratio of 1:5 is most appropriate in our case. The result shows that negative samples are helpful to learn useful news across various topics.

*Embedding size*: The original BERT embedding size is 768, which is too large particularly when we want to use word-level embeddings only. When we use the original embedding vector of BERT, the model performance drops. So, we reduce the embedding size starting from size 50 till 300. We find that the model performance increases till the embedding size of 300. Going forward, the model again drops with larger dimension size (greater than 300). This shows that too small (<50) or too large dimension (>300) reduces the model performance in our experiments. A size of around 300 is best for an NRS to represent the semantic information from the headline and snippet

*Sequence length*: We also exploit the sequence length of reader's clicked history and we find out that too small or too large sequence length results in poor model performance. Too large sequence length results in readers short-term interests being ignored, whereas, too small sequence length will not be able to capture the patterns from readers historical records. Therefore, we choose an average sequence length for training in our model. we test the varying sequence length of a reader's clicks starting from 10 to 300 and find that a length of ~100 reasonably reflects the reader' interests in the news recommendation problem.

*Miscellaneous*: we also test different RNN variants (e.g., GRU versus LSTM), the number of LSTM gates, with or without the layer normalization and the dropout layer, the number of filters and the context window size in the CNN layer, the batch size, and the number of layers. With all these experiments, we find that our current setup is the best for achieving our goals for an NRS.

## 6   Conclusion and Future Work

In this paper, we propose a deep neural network to provide timely, highly accurate and reasonably diverse news recommendations to predict the reader's next click. We learn the news representations by incorporating multiple features from the news. We learn the reader's long-term interests from the whole click history, and the short-term interests and the diversified interests from the recent clicks. We trade-off between high accuracy and reasonable diversity. We apply different attention levels to learn useful news and reader representations. We conduct extensive experiments on two news datasets to demonstrate our work. In the future, we would like to include more readers' feedbacks and address issues such as missing negative implicit feedbacks in NRS. We also like to include position encoding techniques to address the temporal order of news articles and reader clicks. We plan to conduct a user study on whether our proposed method can indeed improve the measurement like click through rate.

## Acknowledgements

This work is partially sponsored by Natural Science and Engineering Research Council of Canada (grant 2020-04760).